# Interplay of 4f-3d Magnetism and Ferroelectricity in DyFeO$_3$

B. Rajeswaran[1], D. Sanyal[2], Mahuya Chakrabarti[3], Y. Sundarayya[1], A. Sundaresan[1] and C.N.R. Rao[1]

[1] *Chemistry and Physics of Materials Unit, Jawaharlal Nehru Centre for Advanced Scientific Research, Jakkur P. O., Bangalore 560 064, India.*
[2] *Variable Energy Cyclotron Centre, 1/AF Bidhannagar, Kolkata 700 064, India*
[3] *Department of Physics, University of Calcutta, 92 APC Road, Kolkata-700009, India*

**Abstract** – DyFeO$_3$ exhibits a weak ferromagnetism ($T_N^{Fe}$ ~ 645 K) that disappears below a spin-reorientation (Morin) transition at $T_{SR}^{Fe}$ ~ 50 K. It is also known that applied magnetic field induces ferroelectricity at the magnetic ordering temperature of Dy-ions ($T_N^{Dy}$ ~ 4.5 K). Here, we show that the ferroelectricity exists in the weak ferromagnetic state ($T_{SR}^{Fe}$ < T < $T_{N,C}$) without applying magnetic field, indicating the crucial role of weak ferromagnetism in inducing ferroelectricity. $^{57}$Fe Mössbauer studies show that hyperfine field ($B_{hf}$) deviates from mean field-like behaviour that is observed in the weak ferromagnetic state and decreases below the onset of spin-reorientation transition (80 K), implying that the $B_{hf}$ above $T_{SR}$ had additional contribution from Dy-ions due to induced magnetization by the weak ferromagnetic moment of Fe-sublattice and below $T_{SR}$, this contribution decreases due to collinear ordering of Fe-sublattice. These results clearly demonstrate the presence of magnetic interactions between Dy(4*f*) and Fe(3*d*) and their correlation with ferroelectricity in the weak ferromagnetic state of DyFeO$_3$.

**Introduction.** – The discovery of ferroelectricity at high magnetic ordering temperature of Fe ($T_N^{Fe}$= 670 K) in one of the rare-earth orthoferrites, SmFeO$_3$ has stimulated the curiosity of researchers working in the field of multiferroics because the rare-earth orthoferrites, RFeO$_3$ (R = rare-earth) crystallize in orthorhombic structure with the centrosymmetric space group *Pbnm (Pnma)* which does not allow spontaneous electric polarization [1-5]. Further, the observed magnetic point symmetry below $T_N^{Fe}$ is also incompatible with the electric polarization [2,3]. However, the magnetic symmetry associated with rare-earth ordering at low temperatures ($T_N^R$) allows existence of linear magnetoelectric effect [4]. In fact, it has been shown experimentally that both GdFeO$_3$ and DyFeO$_3$ exhibit ferroelectric polarization at Neel temperature of Gd and Dy ($T_N^{Gd}$ = 3.2 K and $T_N^{Dy}$ = 4.5 K), but the latter requires an applied magnetic field [6-8]. Meanwhile, a more general observation of ferroelectricity has been reported at the magnetic ordering temperature of chromium ($T_N^{Cr}$ = 50 - 250 K) in isostructural rare-earth orthochromites, RCrO$_3$ with the condition that the R-ion is magnetic and the Cr-sublattice remains weakly ferromagnetic [9]. The origin of ferroelectricity in these materials has been explained by the poling electric field that displaces the magnetic R-ion from its symmetric position which is stabilized by the exchange field on R-ion due to the weak ferromagnetism of Cr-sublattice. This is consistent with the fact that the orthochromites and orthoferrites with nonmagnetic rare-earth ion do not exhibit ferroelectric polarization.

In RFe(Cr)O$_3$, when R-ions are magnetic, there exists three kinds of magnetic interactions, namely Fe(Cr) – O – Fe(Cr), R – O – Fe(Cr) and R – O – R which result in a complex magnetic behaviour [9-11]. A weak interaction between the R – O – R leads to ordering of R ions at low temperatures ($T_N^R$ < 10 K). On the other hand, the R – O – Fe(Cr) interactions result in several interesting properties such as spin-reorientation and magnetization reversal below a compensation temperature

[11-16]. It was shown theoretically that the anisotropic part of magnetic interactions between Fe and R ions are comparable to isotropic Fe – O - Fe interactions and are responsible for spin-reorientation where the direction of easy axis of magnetization of iron sublattice changes its direction from one crystallographic axis to another upon varying the temperature, magnetic field and pressure [10,17]. In some rare-earth orthoferrites, the total magnetization becomes zero at a compensation temperature due to antiferromagnetic coupling of the weak ferromagnetic moment with the $R^{3+}$ ions moment [18].

Here, we report that $DyFeO_3$ exhibits ferroelectricity above $T_{SR}^{Fe}$ ~ 50 K without applying magnetic field and demonstrate the presence of induced magnetization on Dy ions in the weakly ferromagnetic and ferroelectric state and its absence below $T_{SR}^{Fe}$. Our results support the observation of ferroelectricity at the magnetic ordering of Fe in $SmFeO_3$ [1] and consistent with the suggestion that the magnetic interactions between R-Fe(Cr) is responsible for the origin of ferroelectricity in orthoferrites and orthochromites at the Fe(Cr) ordering temperatures, respectively [9].

**Experimental.** – $DyFeO_3$ was prepared by the solid state reaction of stoichiometric quantities of $Dy_2O_3$ and $Fe_2O_3$ at 1673 K for 12 hours followed by several intermittent grinding and heating. Phase purity was confirmed by Rietveld refinement on the powder X-ray diffraction data collected with Bruker D8 Advance diffractometer. Magnetic measurements were carried out with a vibrating sample magnetometer in the magnetic property measurement system (MPMS-SQUID VSM) of Quantum Design, USA. Mössbauer spectra were recorded in transmission mode using $^{57}Co$ γ-ray source in a Rhodium matrix and multi-channel analyzer. Typically more than $10^6$ counts were accumulated in each channel and the spectrometer was calibrated using α-Fe foil. Measurements at low temperatures were carried out using a liquid helium cooled bath cryostat system attached to the sample chamber. The temperature stability at the sample chamber is better than ± 0.2 K. Capacitance and pyroelectric measurements were carried out with LCR meter Agilent E4980A and 6517A Keithley electrometric resistance meter, respectively, using multifunction probe in physical property measurement system (PPMS). Positive-Up and Negative-Down (PUND) measurements were carried out by using Radiant Technologies Inc., precision workstation.

**Results and discussion.** – X-ray diffraction pattern confirmed that the sample crystallizes in orthorhombic perovskite structure with space group *Pbnm*. The lattice parameters and cell volume obtained from profile matching of the X-ray diffraction patterns were consistent with those reported earlier [6]. $DyFeO_3$ is known to undergo antiferromagnetic transition with a weak ferromagnetism at $T_N^{Fe}$ = 645 K, a spin reorientation (Morin) transition at $T_{SR}$ ~ 37 K below which the weak ferromagnetism disappears and at low temperatures the Dy-moments order antiferromagnetically at $T_N^{Dy}$ ~ 4 K [6,8,19]. Due to limitations of accessible temperature range in our PPMS, we have focused both magnetization and pyroelectric measurements in the temperature range 3 – 100 K, which is sufficient to address the problem of our interest. The results of our magnetization measurements, as shown in Fig. 1(a), confirm the presence of spin reorientation transition of Fe-sublattice and magnetic ordering of Dy-ions. It should be noted that the spin-reorientation transition occurs over a wide temperature range with the onset at around 80 K as inferred from magnetization data, as shown in Fig. 1(a), probably due to the polycrystalline nature of the sample. It is difficult to extract the exact temperature of $T_{SR}$ from the magnetization data because of field dependent paramagnetic moment of Dy-ions. Earlier measurements on single crystalline sample showed a spin-reorientation transition of 37 K [7].

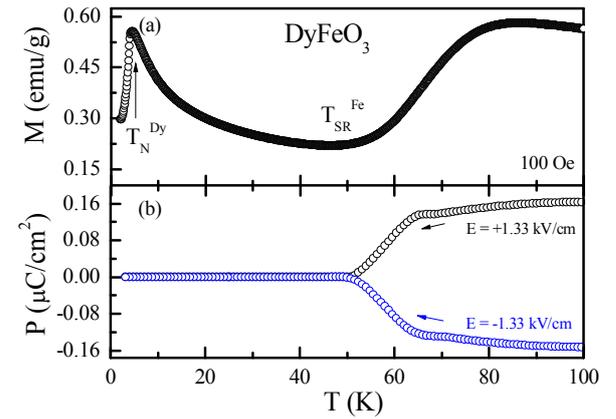

Fig. 1: (Color online) Temperature dependence of (a) Field cooled magnetization of $DyFeO_3$ and (b) Electric polarization near $T_{SR}^{Fe}$.

The electric polarization obtained from pyroelectric measurements in the same temperature range is shown in Fig. 1(b). At first, the sample was poled by applying an electric field of +1.33 kV/cm or -1.33 kV/cm at the maximum temperature achievable in the PPMS (390 K), which is, however, less than the Neel temperature of Fe ordering ($T_N^{Fe}$ ~ 645 K), and then the sample was cooled down to a low temperature (100 K) but higher than $T_{SR}^{Fe}$ (~ 37 K) in the presence of the applied electric field. After shorting the circuit for a reasonably long duration, the current was measured across the spin-reorientation transition using the electrometer down to 3 K at a rate of 4 K/min. Upon integrating the pyrocurrent with respect to time and dividing it by the area of the sample, we obtain electric polarization, which after correcting for leakage is shown in the Fig. 1(b). First of all, there exists ferroelectric polarization much above Dy-ordering temperature in $DyFeO_3$ which is consistent with the observation of electric polarization at high temperature ($T_N^{Fe}$ ~ 670 K) in $SmFeO_3$ [1]. Also, based on the suggestion that the magnetic interactions between rare-earth and Cr(Fe) induce ferroelectric polarization at Neel temperature of Cr(Fe)ordering, we believe that the ferroelectric polarization develops in $DyFeO_3$ at $T_N^{Fe}$ ~ 645 K. Due to ex-perimental limitations and significant leakage at high temperature we could not perform pyroelectric measurement in the vicinity of $T_N$. The important observation in the present study is that the ferroelectric polarization disappears at $T_{SR}$ below which the weak ferromagnetism of Fe-sublattice also disappears, demonstrating that the weak ferromagnetism, in addition to magnetic R-ion (Dy), is indeed an essential component for inducing ferroelectric polarization in $RFeO_3$ which is in agreement with that reported for $ErCrO_3$ [9]. It is also clear from this figure that Dy-ordering ($T_N^{Dy} \sim 4$) alone cannot cause ferroelectricity though the magnetic point symmetry allows linear magnetoelectric effect, in accordance with the report that the applied magnetic field induces weak ferromagnetism and causes ferroelectric polarization at Dy-ordering [6].

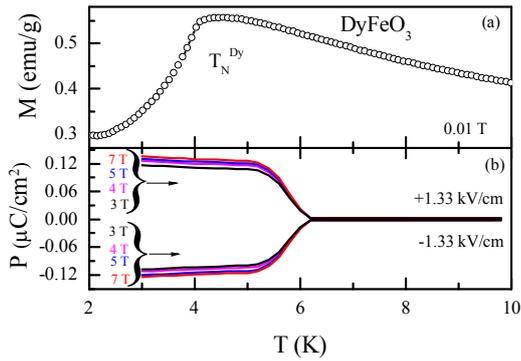

Fig. 2: (Color online) Temperature dependence of (a) Field cooled magnetization of $DyFeO_3$ and (b) Polarization at various magnetic fields near $T_N^{Dy}$.

To measure ferroelectric polarization around Dy ordering temperature ($T_N^{Dy}$), the sample was cooled from 10 K to 3 K in the presence of applied electric field and then the sample was shorted for a long time to minimize the effects to leakage and stray currents. The pyroelectric current is measured upon warming the sample from 3 K to above $T_N^{Dy}$ in the presence of different magnetic fields and the results are shown in Fig. 2. It can be seen from this figure that the polarization remains above $T_N^{Dy}$ measured from magnetization and the value of electric polarization increases with applied magnetic field which is consistent with the fact that the weak ferromagnetic moment increase with field. It also appears from this figure that the electric polarization disappears at above $T_N^{Dy}$. In fact, with induced weak ferromagnetism below $T_{SR}$ by an applied magnetic field, the polarization remains up to $T_N^{Fe}$. Understanding of this result is not straight forward as the pyroelectric current peak leading to polarization appears due to Dy-ordering and not indicative of ferroelectric-paraelectric transition. In general, in such system exhibiting complex magnetic behaviour that controls the electric polarization, case specific poling procedures have to be employed to find the appearance or disappearance of electric polarization [20].

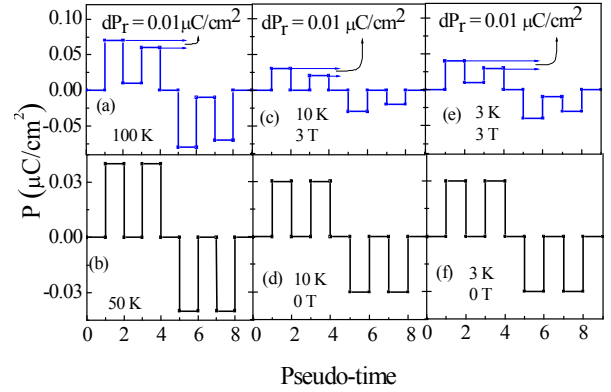

Fig. 3: (Color online) Results of PUND measurement on $DyFeO_3$ at various temperatures. The pulse width is 1000 ms and the pulse delay is maintained at 10 ms throughout the measurements.

However, the PUND measurements are capable of probing the presence or absence of ferroelectricity irrespective of spin-reorientation and Dy-ordering. The results of PUND measurements at various temperatures below 100 K are shown in Fig. 3. The presence of remnant polarization at 10 K under an applied magnetic field (3 T) confirms the ferroelectricity below $T_{SR}$. It should be noticed that there is no remnant polarization in the absence of magnetic field. It is also to be mentioned that the spin reorientation transition can be suppressed by cooling the sample in a magnetic field ($\pm$ 3 T) and/or electric field and the weak ferromagnetism induced ferroelectric polarization can be stabilized down to 3 K.

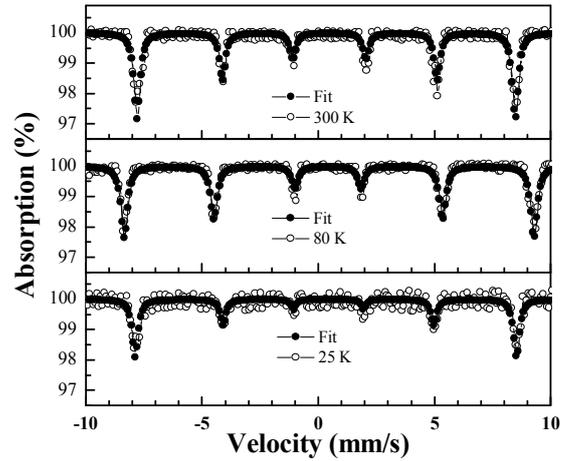

Fig. 4: (Color online) Mössbauer spectra of $DyFeO_3$ at 300, 80 and 25 K.

'Zero-field' $^{57}Fe$ Mössbauer spectra recorded on $DyFeO_3$ at different temperatures (T < 300 K) showed six line patterns confirming the magnetic ordering of Fe-sublattice as reported earlier.[21]

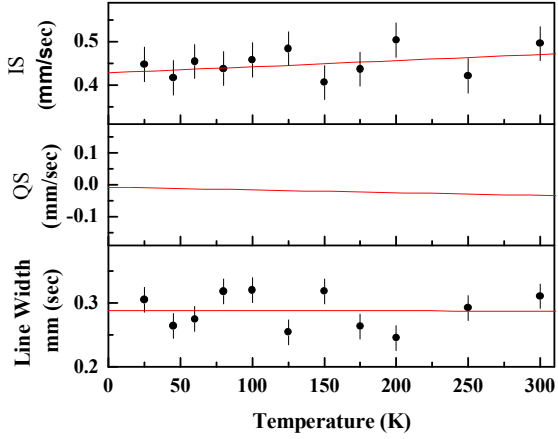

Fig. 5: (Color online) Temperature dependence of hyperfine parameters such as isomer shift, quadrupole splitting and linewidth in DyFeO$_3$. The line passing through the data is the linear fit.

In the P*bnm* structure, since the Fe-ions occupy the only one crystallographic site (4*b*), we expect that the Mössbauer spectra should contain only a single sextet. Examination of linewidth of all the spectra gave a value of ~ 0.3 mm/s which is typical of that observed for magnetic spectrum. Consequently, each spectrum was fitted with six independent Lorentizian shaped lines and the results at 300, 80 and 25 K are shown in Fig. 4. The obtained hyperfine parameters such as isomer shift, quadrupole splitting and line width at various temperatures are shown in Fig. 5. It can be seen that all these parameters are almost independent of temperature and the value of isomer shift is consistent with trivalent state of iron. The obtained hyperfine interaction parameters are consistent with the earlier reported values for DyFeO$_3$[21]. However, the earlier studies were focused mainly at high temperatures (T ≥ 85 K) but an extrapolated value of $B_{hf}$ = 548 kOe at 0 K was given that is consistent with our result.

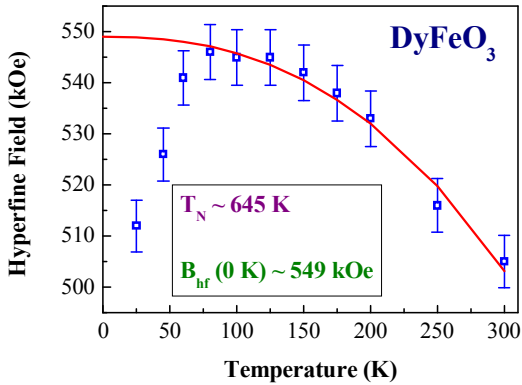

Fig. 6: (Color online) Temperature dependence of $B_{hf}$ in the range 25 – 300 K. The decrease of $B_{hf}$ at 80 K shows decoupling of Dy-moments from Fe-sublattice.

The most important observation in the present study is the behaviour of temperature dependence of the hyperfine field ($B_{hf}$) which deviates from the mean field-like behaviour observed in the weak ferromagnetic state and decreases below the onset of spin-reorientation temperature (80 K) as shown in Fig. 6. Though the $B_{hf}$(T) in the weak ferromagnetic state appears to follow mean field behaviour, due to limited range of temperature we have used an empirical relationship, ($B_{hf}$(T)/$B_{hf}$(0) = B(1-T/T$_N$)$^\beta$ to obtain the extrapolated value at low temperatures [22]. It can be seen that the aforementioned expression fits the experimental data fairly well. It should be noted that the value of $B_{hf}$ at 25 K is 40 kOe less than the corresponding extrapolated value in the weak ferromagnetic state. Such a decrease of $B_{hf}$ below the Morin transition ($T_{SR}$) is not known earlier. Moreover, the decrease of $B_{hf}$ cannot be attributed to weak ferromagnetic to collinear transition of Fe-sublattice alone because a similar spin-reorientation (Morin-type) transition in YFe$_{0.85}$Mn$_{0.15}$O$_3$ with nonmagnetic Y-ion does not exhibit such a decrease in $B_{hf}$ below $T_{SR}$.[23] The relatively high value of $B_{hf}$ in the weak ferromagnetic state indicates that the iron nucleus experiences a local field that results from combined effect of Fe and Dy neighbours. We suggest that the local field at Dy-ions is induced by the weak ferromagnetic moment of Fe-sublattice. If Dy-ions were simple paramagnetic, we would not see such a large decrease (40 kOe) of $B_{hf}$ below $T_{SR}$. When the weak ferromagnetism of iron sublattice disappears below $T_{SR}$, the contribution from Dy-moments and thus the ferroelectricity also disappear. The observation that the $B_{hf}$ decreases below $T_{SR}$ suggest that the weak ferromagnetism and magnetic R-ion are essential for ferroelectricity at $T_N^{Fe}$. In fact, electronic structure analysis of DyFeO$_3$ has shown that there is a coupling between Dy and Fe-ions which occurs through hybridized Dy-d and O-2p states [8]. It has been suggested that the electric polarization at $T_N^{Cr}$ in isostructural RCrO$_3$ occurs due to combined effect of electric poling and induced exchange striction [9]. We propose that a similar mechanism operates in DyFeO$_3$.

In conclusion, our study on DyFeO$_3$ reveals that ferroelectricity exists in the weak ferromagnetic state and disappears below spin-reorientation (Morin) transition without applying a magnetic field, thereby establishing the crucial role of weak ferromagnetism in inducing ferroelectricity. Temperature dependence of hyperfine field from Mossbauer studies show that there is an induced magnetization on Dy-sublattice by the weak ferromagnetic moment of Fe-sublattice and the induced magnetization decreases below the onset of Morin transition. These results clearly establish the interplay of R(4f) – Fe(3d) magnetism and ferroelectricity. Our study further suggests that the magnetic ordering of R-ions is not essential for inducing ferroelectricity but a paramagnetic R-ion and weak ferromagnetism of Fe-ions are sufficient.

***

One of the authors (A. S) would like to thank Sheikh Saqr Laboratory at International Centre for Materials Science,

Jawaharlal Nehru Centre for Advanced Scientific Research, Bangalore and also Department of Science and Technology for funding under the joint Indo-Japan project. One of the authors (MC) acknowledges UGC, Government of India for providing financial assistance as Dr. D.S. Kothari post-doctoral fellowship.